\documentclass[aps,prl,reprint,twocolumn,amsmath,amssymb,citeautoscript]{revtex4-1}
\usepackage{graphicx}
\usepackage{dcolumn}
\usepackage{bm}
\usepackage{latexsym,epsfig}
\usepackage{graphicx}
\usepackage{verbatim}
\usepackage{comment}
\usepackage{amsmath}
\usepackage{amssymb}
\usepackage{physics}
\usepackage{stmaryrd}
\usepackage{color}
\usepackage{epstopdf}
\usepackage{grffile}
\usepackage{lipsum}
\usepackage{enumitem}
\usepackage[normalem]{ulem}
\DeclareGraphicsExtensions{.eps}

\newcommand{\beq}{\begin{equation}}
	\newcommand{\eeq}{\end{equation}}
\newcommand{\bea}{\begin{eqnarray}}
	\newcommand{\eea}{\end{eqnarray}}
\newcommand{\ben}{\begin{eqnarray*}}
	\newcommand{\een}{\end{eqnarray*}}
\newcommand{\bfig}{\begin{figure}}
	\newcommand{\efig}{\end{figure}}

\usepackage{mathtools}
\usepackage{braket}
\usepackage[dvipsnames]{xcolor}
\usepackage{float}
\usepackage{subfigure}
\usepackage{upgreek}
\usepackage{pgffor}
\usepackage{silence}
\usepackage[colorlinks=true,linktoc=page,linkcolor=blue,citecolor=magenta,urlcolor=Fuchsia]{hyperref}
\usepackage[normalem]{ulem}
\usepackage{sidecap,tikz}
\usepackage{orcidlink}
\usepackage{makecell}
\usepackage{mathrsfs}
\usepackage{comment}

\mathchardef\mhyphen="2D

\begin{document}
	
	\title{Spin-resolved spectral topology and re-entrant localization in a non-Hermitian quasiperiodic SSH chain}
	
	\author{Hemant K Sharma}
	\email{hemant214786@gmail.com}
	\affiliation{Department of Physics, National Institute of Science Education and Research, Jatni, 752050, India}
	\affiliation{Homi Bhabha National Institute, Training School Complex, Anushakti Nagar, Mumbai 400094, India}
	
	\date{\today}
	
	\begin{abstract}
		
		We investigate localization and spectral topology in a non-Hermitian quasiperiodic Su--Schrieffer--Heeger lattice with Rashba spin--orbit coupling and spin-dependent hopping. By analyzing the inverse participation ratio, complex-energy spectrum, and spectral winding numbers, we demonstrate the emergence of a re-entrant transition from extended to localized and back to extended phases as the non-Hermitian parameter increases. The localization transition is accompanied by a simultaneous real-complex-real spectral transition in the complex-energy plane. In the absence of spin-dependent hopping, the spectrum forms two nearly spin-degenerate loops characterized by winding numbers $w=\pm2$. Upon introducing finite spin-dependent hopping, each loop splits into two independent spin-resolved spectral branches, resulting in four disconnected spectral contours carrying distinct winding sectors. Our results reveal a direct correspondence between localization, spectral topology, and spin-resolved spectral splitting in non-Hermitian quasiperiodic systems.
		
	\end{abstract}

	\maketitle 
	\section{Introduction}
	
	Quasiperiodic systems provide an important platform for studying localization phenomena in low-dimensional quantum systems, lying intermediate between perfectly periodic crystals and completely disordered media~\cite{DominguezCastro2019}. Unlike random one-dimensional systems, where arbitrarily weak disorder localizes all eigenstates, quasiperiodic lattices exhibit sharp localization transitions at finite modulation strength. A paradigmatic example is the Aubry--Andr\'e--Harper (AAH) model, which undergoes a delocalization--localization transition driven by a quasiperiodic onsite potential~\cite{AubryAndre1980,Harper1955}. In addition to localization physics, the AAH model has also attracted considerable attention due to its close connection with topological phases and quantum Hall physics.
	
	Generalizations of the AAH model have revealed a variety of rich phenomena, including mobility edges, critical phases, and the coexistence of localized and extended states within the same spectrum~\cite{Ganeshan2015,Li2020,Li2017,Luschen2018,Biddle2010}. More recently, non-Hermitian extensions of quasiperiodic lattices have emerged as an active area of research due to the appearance of fundamentally new effects such as complex energy spectra, exceptional points, non-Hermitian skin effects, and spectral winding topology~\cite{Zeng2020,Zeng2020PRR,Zeng2017,Claes2021,Longhi2021,Jiang2019,Cai2021a,Tang2021,Yao2018,Kawabata2019}. In many such systems, the localization transition is closely connected to a real-to-complex spectral transition and the emergence of finite spectral winding numbers~\cite{Longhi2019PRL,Liu2021b,Liu2020,Chen2022,Lin2022,ZhouHan2022,Cai2021b,ZhouHan2021,Acharya2022,Zhou2021,Xu2021}. These studies have established a direct correspondence between localization and spectral topology in non-Hermitian quasiperiodic systems.
	
	Parallel to these developments, the Su--Schrieffer--Heeger (SSH) model has become a prototypical framework for understanding one-dimensional topological phases through dimerized hopping and bulk--edge correspondence~\cite{Su1979,Heeger1988,Asboth2016}. The interplay between quasiperiodicity and SSH topology has been shown to produce topological mobility edges, critical states, and quasiperiodicity-driven topological transitions~\cite{Ganeshan2015,Biddle2010,Liu2020b,Xu2021}. Furthermore, non-Hermitian generalizations of the SSH model exhibit unconventional topological phases, spectral winding, and complex-energy loop structures~\cite{Yao2018,Longhi2019PRB,Longhi2019PRL,Kawabata2019,Gong2018,ElGanainy2018}.
	
	Despite these advances, the combined role of quasiperiodicity, Rashba spin--orbit interaction, and spin-dependent hopping in determining the localization and spectral topology of non-Hermitian SSH systems remains largely unexplored. In particular, the effect of spin-dependent hopping on the formation of spin-resolved spectral loops and winding-number splitting has not yet been systematically investigated.
	
	In this work, we study a quasiperiodic SSH chain with Rashba spin--orbit coupling and spin-dependent hopping in the presence of a complex quasiperiodic onsite potential. The model realizes a generalized non-Hermitian SSH--AAH lattice exhibiting rich localization and topological behavior. By analyzing the inverse participation ratio (IPR), complex-energy spectrum, and spectral winding numbers, we demonstrate the emergence of a re-entrant delocalization--localization--delocalization transition as the non-Hermitian parameter $h$ is varied. Simultaneously, the system undergoes a real--complex--real spectral transition accompanied by the appearance and collapse of nontrivial spectral loops in the complex-energy plane.
	
	A central result of the present work is the emergence of spin-resolved spectral topology induced by the spin-dependent hopping term $g_h$. In the absence of spin-dependent hopping, the complex-energy spectrum consists of two nearly spin-degenerate spectral loops characterized by winding numbers $w=\pm2$. Upon introducing finite $g_h$, each spectral contour splits into two independent spin-polarized branches, increasing the number of disconnected loops from two to four. Consequently, multiple winding sectors emerge, reflecting the lifting of spin degeneracy and the formation of spin-resolved topological branches in the complex-energy plane.
	
	We further show that the appearance of finite winding numbers coincides with enhanced localization and complex spectral formation, establishing a direct connection between localization, spectral topology, and spin-resolved spectral splitting. At larger values of the non-Hermitian parameter, the spectral loops collapse back onto the real axis, the winding numbers vanish, and the system re-enters a topologically trivial extended phase. The resulting sequence therefore realizes a re-entrant interplay between localization, spectral topology, and non-Hermitian symmetry breaking.
	
	Our results demonstrate how spin-dependent hopping can fundamentally reshape the spectral topology of non-Hermitian quasiperiodic lattices and generate multiple spin-resolved winding sectors. The present model provides a promising platform for exploring controllable non-Hermitian topological phases and spin-selective localization phenomena in ultracold atoms, photonic lattices, topolectric circuits, and other synthetic quantum systems.
	
	{\em Model.-} 
	The non-Hermitian generalized AAH model is defined as 
	\begin{align}
		H = &\sum_{n=1}^{L} \Big( 
		v\, c_{n}^{B\dagger} c_{n}^{A} 
		+ \tilde{w}\, c_{n+1}^{A\dagger} c_{n}^{B} 
		+ \text{H.c.} 
		\Big) \nonumber\\
		&+ \lambda \sum_{n=1}^{L} 
		\frac{\cos(2\pi\beta n + \phi)}{1 - \alpha \cos(2\pi\beta n + \phi)} 
		\Big(c_{n}^{A\dagger} c_{n}^{A} -c_{n}^{B\dagger} c_{n}^{B}\Big)\nonumber\\
		\label{eq:ham}
	\end{align}
	where $\tilde{w}= w +i g_h$
	\begin{align}
		H_{R}=  \sum_{n} \; & 	
		i  \, c^{\dagger B}_{n} (\alpha_1\sigma^z+\alpha_3\sigma^y) c^{A}_{n}
		+ i  \, c^{\dagger A}_{n+1} (\alpha_2\sigma^z+\alpha_4\sigma^y) c^{B}_{n} \nonumber \\
	\end{align}
	Here, $c_{n}^{A(B)\dagger}$ ($c_{n}^{A(B)}$) denotes the creation (annihilation) operator of a spinless fermion at the $A$ ($B$) sublattice of the $n$th unit cell. The parameters $v$ and $w$ represent the intracell and intercell hopping amplitudes, respectively, while $\lambda$ quantifies the strength of the quasiperiodic onsite potential. The parameter $\beta = (\sqrt{5} - 1)/2$ is an irrational number known as the inverse golden ratio, and $\phi$ denotes the phase of the modulation. Although the hopping amplitude contains an imaginary component \(ig_h\), the hopping sector remains Hermitian because the Hermitian conjugate term is included explicitly.
	
	\par
	Non-Hermiticity is introduced into the system through a complex phase, $\phi = \theta + i h$, where the imaginary component $h$ controls the degree of Hermiticity breaking. The non-Hermitian generalized Aubry--Andr\'e--Harper (nHGAAH) model preserves $\mathcal{PT}$ symmetry when the real part of the phase is set to zero, i.e., $\theta = 0$. Unless otherwise stated, we adopt this condition throughout the paper. 
	
	\par
	The parameters $\alpha_{1}$ and $\alpha_{2}$ correspond to the Rashba spin--orbit interaction (RSOI) terms that preserve spin orientation (spin-conserving processes), whereas $\alpha_{3}$ and $\alpha_{4}$ account for the RSOI terms involving spin-flip processes.

	{\em Results.-}
	For $\alpha = 0$ and $h = 0$, the model reduces to the Hermitian one-dimensional dimerized lattice with staggered quasiperiodic disorder. The interplay between dimerization and quasiperiodic modulation gives rise to a re-entrant localization transition, as previously reported in Ref.~\cite{shilpi}. In the present work, we show that upon introducing non-Hermiticity, the system exhibits a sequence of three distinct phase transitions as a function of the Hermiticity-breaking parameter $h$. Remarkably, at larger values of $h$, the system reverts to its original phase by undergoing another set of triple transitions, thereby realizing a re-entrant topological–localization behavior. These results are obtained by numerically diagonalizing the Hamiltonian in Eq.~\ref{eq:ham}  using exact diagonalization. Throughout this work, we set $v = 1$ and $w = 1.5$ as the energy scale and fix the quasiperiodic potential strength at $\lambda = g_h = 1$.
	\begin{figure}[t]
		\centering
		\includegraphics[width=1\columnwidth]{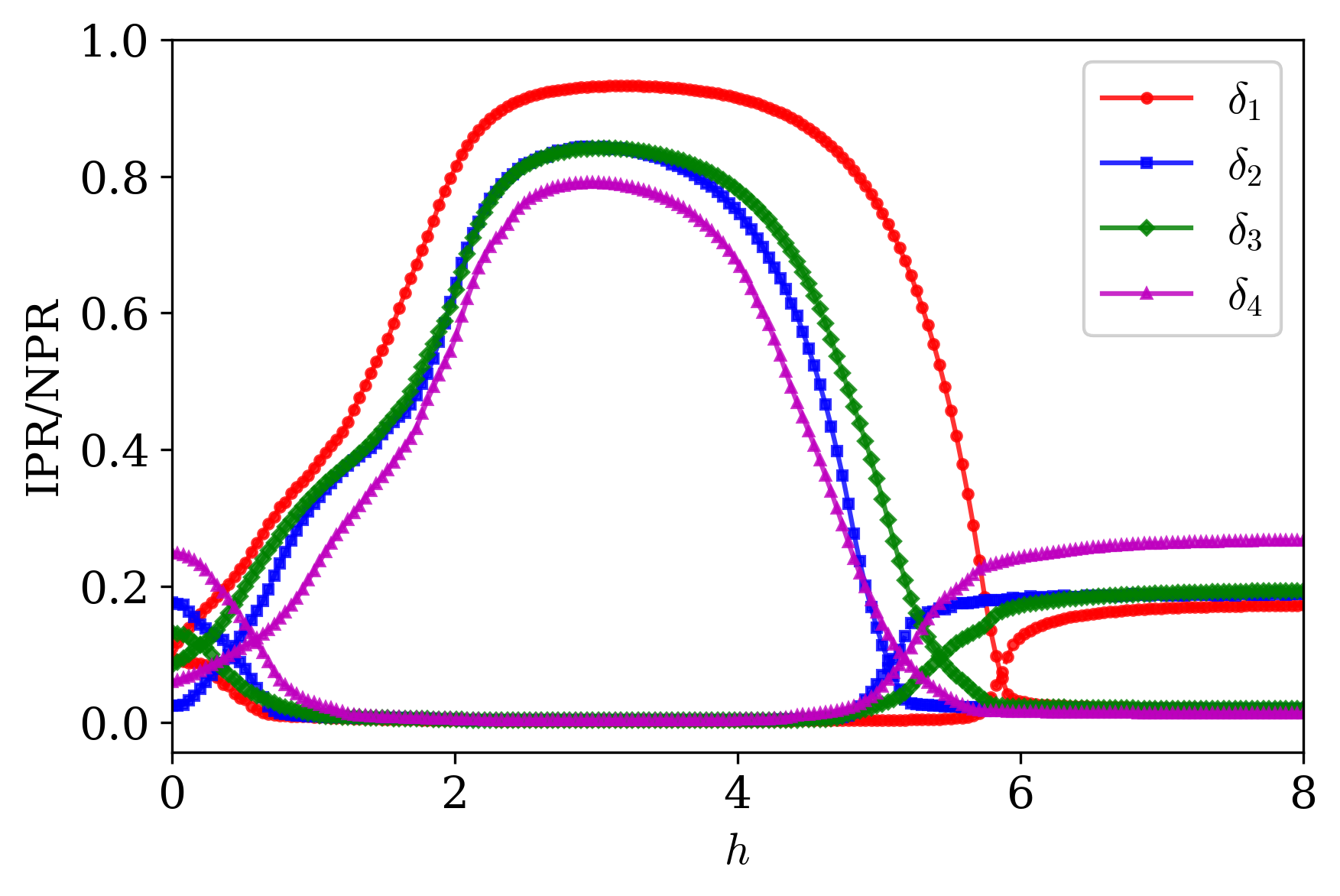}

		\caption{
			Variation of $\langle \mathrm{IPR} \rangle$ and the average normalized participation ratio $\langle \mathrm{NPR} \rangle$ as a function of $h$ for $\alpha = 0.2$, demonstrating a re-entrant localization transition.
			Here $\delta_1$ means all parameter except v ,w h,$\alpha$ are zero $\delta_2 $ include $ \alpha_1 = 1, \alpha2= 2$ $\delta_3 $ further have $\eta=1$ and $\delta_4 $ also include $ \alpha_3 = 1, \alpha_4= 0.5$ L = 1597.
		}
		\label{fig:IPR2}
	\end{figure}

	\begin{figure}[t]
		\centering
		\includegraphics[width=1\columnwidth]{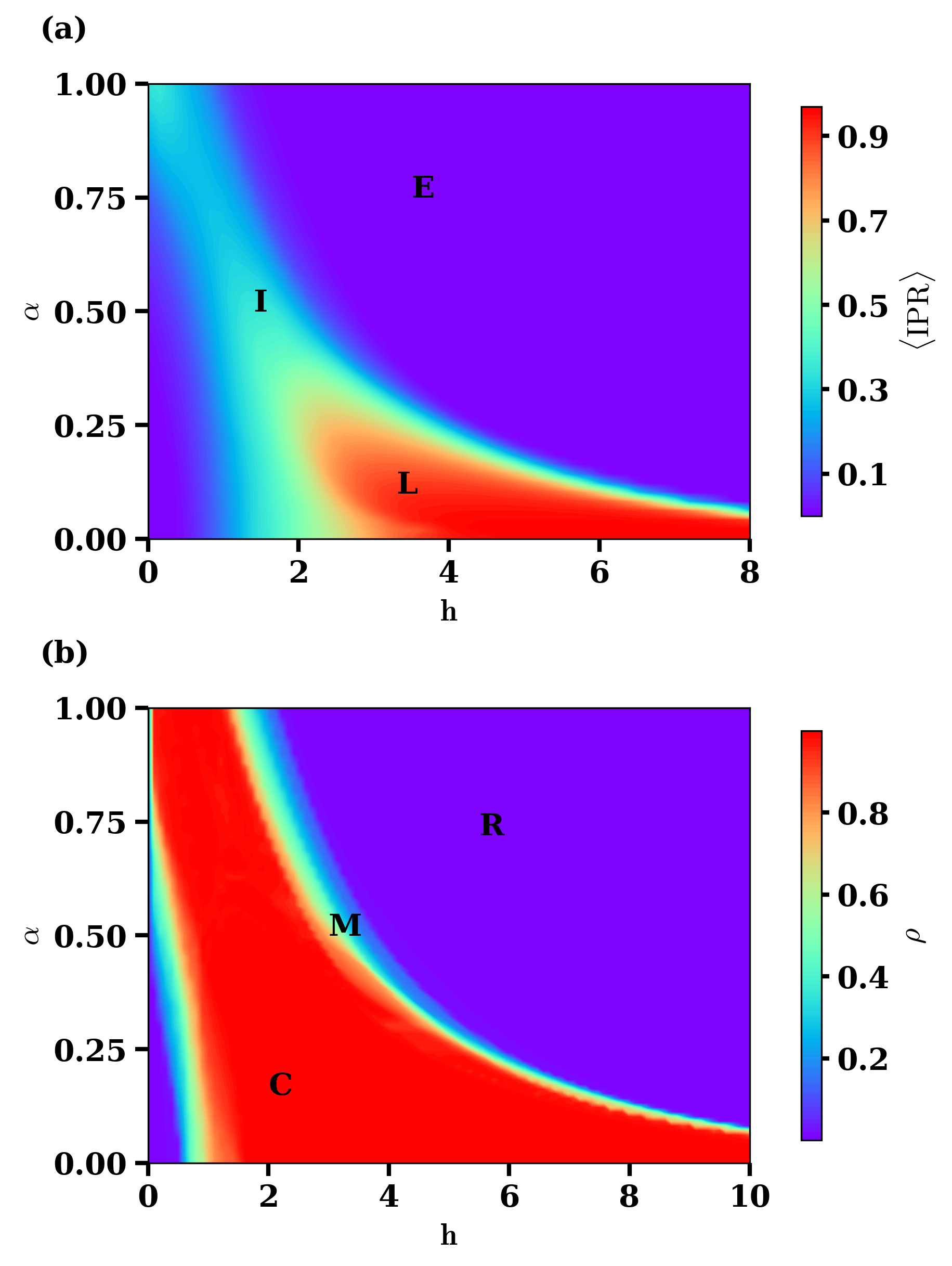}

		\caption{
			(a) Phase diagram in the $\alpha$--$h$ plane obtained from the average inverse participation ratio, $\langle \mathrm{IPR} \rangle$, showing the extended (E), intermediate (I), and localized (L) phases. 
			The color map represents the magnitude of $\langle \mathrm{IPR} \rangle$.
			(b) Phase diagram in the $\alpha$--$h$ plane constructed from the density of states $\rho$, distinguishing the $\mathcal{PT}$-unbroken (R), mixed (C), and $\mathcal{PT}$-broken (B) phases. L = 610
		}
		\label{fig:IPR1}
	\end{figure}
	
	In the following, we discuss our findings in detail. First we will focus on the delocalization-localization transition. Next we will investigate the transition related to the $\mathcal{P}\mathcal{T}$ symmetry breaking and then explore the spectral topological character associated to these transitions. 
	{\em Delocalization--localization transition.-}
	
	To characterize the localization properties of the system, we first compute the average inverse participation ratio (IPR) and normalized participation ratio (NPR). 
	The inverse participation ratio for the $n$th eigenstate is defined as
	\begin{equation}
		\mathrm{IPR}_{n}=\sum_{j=1}^{L}|\psi_{n}^{j}|^{4},
	\end{equation}
	where $\psi_{n}^{j}$ denotes the amplitude of the $n$th eigenstate at lattice site $j$ and $L$ is the system size. 
	For extended states, the wavefunction spreads over the entire lattice and $\mathrm{IPR}_{n}\rightarrow 0$ in the thermodynamic limit, whereas localized states retain finite IPR values. 
	The corresponding normalized participation ratio is given by
	\begin{equation}
		\mathrm{NPR}_{n}=\frac{1}{L\,\mathrm{IPR}_{n}}.
	\end{equation}
	To characterize the overall localization behavior of the spectrum, we evaluate the average quantities
	\begin{equation}
		\langle \mathrm{IPR} \rangle
		=\frac{1}{N}\sum_{n=1}^{N}\mathrm{IPR}_{n},
	\end{equation}
	and similarly $\langle \mathrm{NPR} \rangle$, where $N$ is the total number of eigenstates.
	
	Figure~\ref{fig:IPR2} shows the variation of $\langle \mathrm{IPR} \rangle$ and $\langle \mathrm{NPR} \rangle$ as a function of the non-Hermitian parameter $h$ for different coupling configurations. 
	For small values of $h$, the system remains predominantly extended, characterized by small $\langle \mathrm{IPR} \rangle$ and large $\langle \mathrm{NPR} \rangle$. 
	As $h$ increases, the IPR increases rapidly while the NPR decreases, signaling the onset of localization. 
	At intermediate values of $h$, the system enters a strongly localized regime where $\langle \mathrm{IPR} \rangle$ reaches its maximum and $\langle \mathrm{NPR} \rangle$ becomes minimal. 
	Interestingly, upon further increasing $h$, the localization strength decreases again and the system gradually recovers an extended phase. 
	This demonstrates a clear re-entrant localization transition driven by the non-Hermitian quasiperiodic modulation.
	
	The effect of Rashba spin--orbit interaction and spin-dependent hopping is clearly visible in the different curves shown in Fig.~\ref{fig:IPR2}. 
	The parameter set $\delta_{1}$ corresponds to the case without spin--orbit coupling and spin-dependent hopping. 
	The set $\delta_{2}$ includes only the spin-conserving Rashba couplings $\alpha_{1}=1$ and $\alpha_{2}=2$, while $\delta_{3}$ further incorporates the spin-dependent hopping term $g_{h}=1$. 
	Finally, $\delta_{4}$ additionally includes the spin-flip Rashba terms $\alpha_{3}=1$ and $\alpha_{4}=0.5$. 
	
	The inclusion of spin--orbit coupling shifts the localization boundaries and modifies the width of the intermediate regime. 
	In particular, the spin-dependent hopping and spin-flip Rashba terms suppress the fully localized phase and broaden the coexistence region between localized and extended states. 
	
	To obtain a global picture of the localization transition, we next compute the average inverse participation ratio in the $h$--$\alpha$ plane, as shown in Fig.~\ref{fig:IPR1}(a). 
	The phase diagram reveals three distinct regions identified by the magnitude of $\langle \mathrm{IPR} \rangle$. 
	The region with very small $\langle \mathrm{IPR} \rangle$ corresponds to the extended (E) phase, whereas large values indicate the localized (L) phase. 
	Between these two regimes, an intermediate (I) phase emerges where localized and extended states coexist within the spectrum, indicating the presence of mobility edges.
	
	A particularly important feature of the phase diagram is the re-entrant localization transition induced by the non-Hermitian parameter $h$. 
	Starting from the extended phase at small $h$, increasing $h$ first drives the system into the intermediate regime and subsequently into a fully localized phase. 
	However, upon further increasing $h$, the localized phase gradually shrinks and the system again enters an intermediate regime before finally recovering the extended phase at large $h$. 
	Consequently, the system undergoes the re-entrant sequence
	\begin{equation}
		E \rightarrow I \rightarrow L \rightarrow I \rightarrow E.
	\end{equation}
	
	{\em Real--complex spectral transition.-}
	
	We next analyze the spectral properties of the system through the fraction of eigenstates possessing nonzero imaginary energies. 
	To characterize the transition between real and complex spectra, we define
	\begin{equation}
		\rho=\frac{N_{c}}{L},
	\end{equation}
	where $N_{c}$ denotes the number of eigenvalues with finite imaginary parts. 
	In the thermodynamic limit, $\rho=0$ corresponds to a completely real spectrum, whereas $\rho=1$ indicates that all eigenvalues are complex.
	
	The corresponding phase diagram is shown in Fig.~\ref{fig:IPR1}(b). 
	Three distinct spectral regions can be identified. 
	The region with $\rho=0$ corresponds to the real-spectrum (R) phase where all eigenvalues remain real. 
	In contrast, the region with $\rho=1$ represents the fully complex-spectrum (C) phase in which all eigenvalues acquire finite imaginary components. 
	Between these limits lies a mixed (M) phase characterized by the coexistence of real and complex eigenvalues.
	
	Remarkably, the spectral phase diagram exhibits a re-entrant behavior analogous to the localization transition. 
	At small values of $h$, the spectrum remains completely real. 
	As $h$ increases, complex eigenvalues gradually emerge and the system enters the mixed phase. 
	Further increase in $h$ drives the system into the fully complex regime. 
	However, beyond a critical value of $h$, the number of complex eigenvalues decreases again and the system re-enters the mixed regime before finally recovering a predominantly real spectrum at larger $h$. 
	Thus, the system undergoes the re-entrant spectral transition
	\begin{equation}
		R \rightarrow M \rightarrow C \rightarrow M \rightarrow R.
	\end{equation}
	
	The strong correspondence between the $\langle \mathrm{IPR} \rangle$ and $\rho$ phase diagrams establishes a direct connection between localization and spectral topology in the present quasiperiodic non-Hermitian system. 
	The localized regime is accompanied by the emergence of complex eigenvalues, while the extended phases predominantly possess real spectra. 
	The intermediate regime simultaneously hosts localized and extended states together with coexisting real and complex eigenvalues, revealing the rich interplay between quasiperiodicity, non-Hermiticity, Rashba spin--orbit coupling, and spin-dependent hopping.

	\begin{figure}[t]
		\centering
		\includegraphics[width=1\columnwidth]{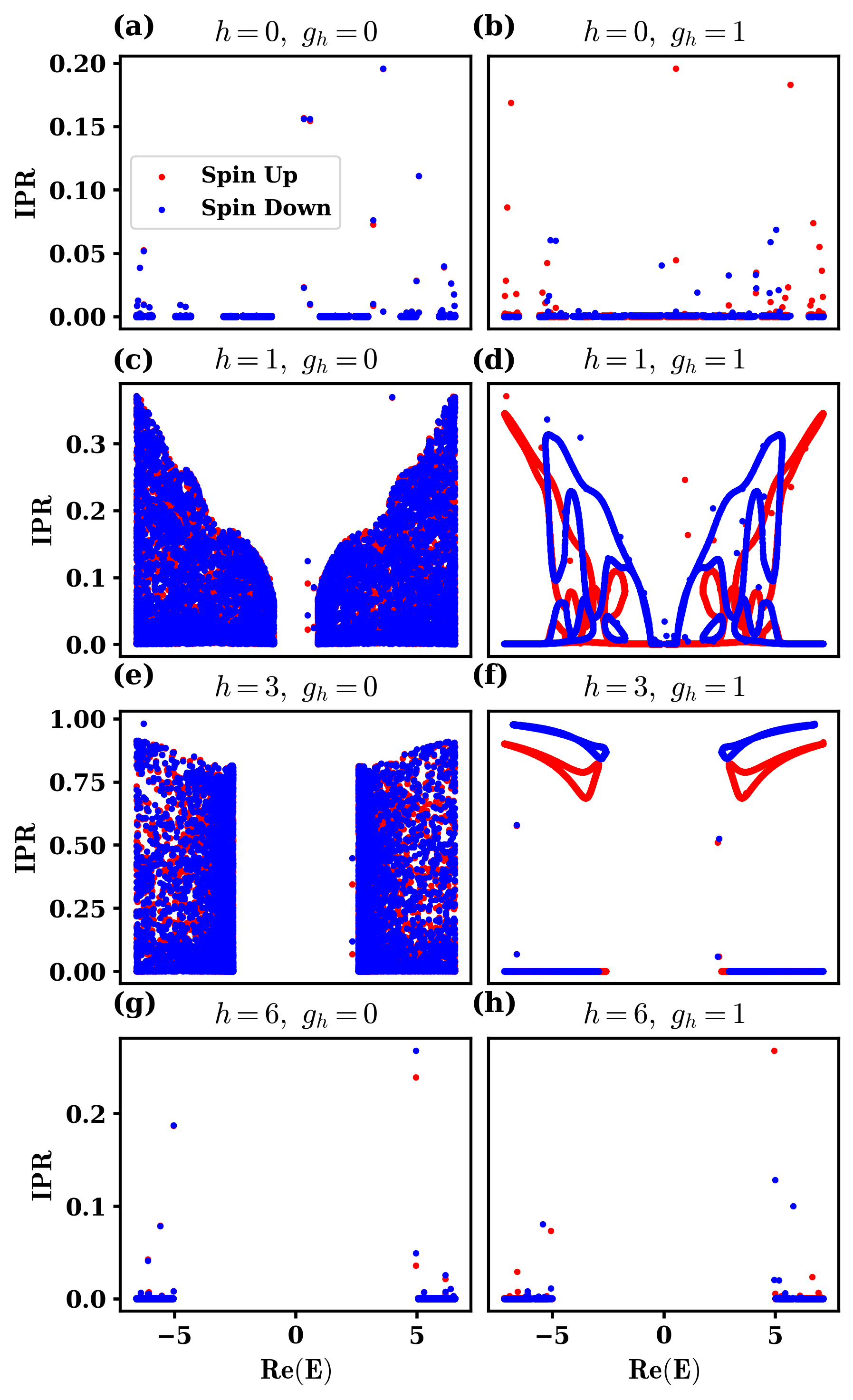}
		\caption{
			Spin-resolved inverse participation ratio (IPR) as a function of $\mathrm{Re}(E)$ for different values of the non-Hermitian parameter $h$ and spin-dependent hopping $g_h$. Red and blue dots correspond to the spin-up and spin-down sectors, respectively.L = 1597
		}	
		\label{fig:spinipr}
	\end{figure}
	
	{\em Spin-resolved inverse participation ratio.-}
	
	To characterize the localization properties of the spin-resolved eigenstates, we compute the inverse participation ratio (IPR) separately for the spin-up and spin-down sectors. 
	The IPR provides a quantitative measure of the spatial localization of an eigenstate. 
	For an extended state, the wavefunction spreads over the entire lattice and the IPR approaches zero in the thermodynamic limit, whereas for a localized state the IPR remains finite.
	
	For the $j$th eigenstate, the spin-resolved inverse participation ratios are defined as
	\begin{equation}
		\mathrm{IPR}^{\uparrow}_{j}
		=
		\frac{
			\sum\limits_{n=1}^{N}
			\left|
			\psi^{\uparrow}_{j,n}
			\right|^{4}
		}{
			\left(
			\langle \psi_j | \psi_j \rangle
			\right)^2
		},
		\label{eq:ipr_up}
	\end{equation}
	
	and
	
	\begin{equation}
		\mathrm{IPR}^{\downarrow}_{j}
		=
		\frac{
			\sum\limits_{n=1}^{N}
			\left|
			\psi^{\downarrow}_{j,n}
			\right|^{4}
		}{
			\left(
			\langle \psi_j | \psi_j \rangle
			\right)^2
		},
		\label{eq:ipr_down}
	\end{equation}
	where $\psi^{\uparrow}_{j,n}$ and $\psi^{\downarrow}_{j,n}$ denote the spin-up and spin-down components of the $j$th eigenstate at lattice site $n$, respectively, and $N$ represents the total number of lattice sites. 
	The denominator ensures proper normalization of the eigenstates.
	
	Using Eqs.~(\ref{eq:ipr_up}) and (\ref{eq:ipr_down}), we analyze the spin-resolved localization properties as a function of the real part of the energy spectrum, $\mathrm{Re}(E)$, for different values of the non-Hermitian parameter $h$ and hopping parameter $g_h$, as shown in Fig.~\ref{fig:spinipr}. 
	The red and blue dots correspond to the spin-up and spin-down sectors, respectively.
	
	For $h=0$ [Figs.~\ref{fig:spinipr}(a) and \ref{fig:spinipr}(b)], almost all eigenstates possess very small IPR values clustered near zero, indicating that the system predominantly remains in an extended phase. 
	Only a few isolated states exhibit slightly larger IPR values, corresponding to weakly localized edge or quasibound states. 
	The spin-up and spin-down sectors almost overlap completely, suggesting that the localization properties remain nearly spin symmetric in the weakly non-Hermitian regime.
	
	As the non-Hermitian parameter increases to $h=1$, the localization properties change significantly. 
	For $g_h=0$ [Fig.~\ref{fig:spinipr}(c)], a broad distribution of finite IPR values emerges over a large energy window, signaling the onset of localization. 
	The coexistence of low-IPR and high-IPR states demonstrates the emergence of mobility edges separating extended and localized eigenstates. 
	The localization becomes stronger near the outer spectral branches, while states around the central energy region remain comparatively extended.
	
	For finite hopping modulation $g_h=1$ [Fig.~\ref{fig:spinipr}(d)], the localization landscape becomes highly structured. 
	The eigenstates organize into several arc-like branches in the $(\mathrm{Re}(E), \mathrm{IPR})$ plane, reflecting strong spectral reorganization induced by the hopping modulation. 
	A clear asymmetry between the spin-up and spin-down sectors becomes visible, particularly near the outer spectral branches where the spin-down states acquire larger IPR values, indicating enhanced spin-selective localization.
	
	At $h=3$ [Figs.~\ref{fig:spinipr}(e) and \ref{fig:spinipr}(f)], most eigenstates exhibit very large IPR values concentrated near the outer spectral branches, implying that the system predominantly enters a localized phase. 
	The coexistence of highly localized states together with states possessing nearly vanishing IPR values again indicates mobility-edge-like behavior. 
	For $g_h=1$, the localization structure becomes smoother and more organized, with the spin-up and spin-down sectors forming distinct spectral branches characterized by different localization strengths.
	
	Finally, for large non-Hermitian strength $h=6$ [Figs.~\ref{fig:spinipr}(g) and \ref{fig:spinipr}(h)], most eigenstates collapse back to nearly zero IPR values, indicating a recovery of the extended phase. 
	Only a few isolated states near the spectral edges retain finite IPR values corresponding to residual localized states. 
	This behavior demonstrates the re-entrant localization-delocalization transition induced by the interplay of quasiperiodicity, spin-orbit interaction, and non-Hermiticity.
	
	Overall, the spin-resolved IPR analysis reveals the emergence of mobility edges, spin-selective localization, and re-entrant localization transitions in the non-Hermitian quasiperiodic system. 
	The hopping modulation parameter $g_h$ significantly reshapes the localization landscape and enhances the separation between the spin-up and spin-down localization properties.\\

	\begin{figure*}[t]
		\centering
		\includegraphics[width=0.98\textwidth]{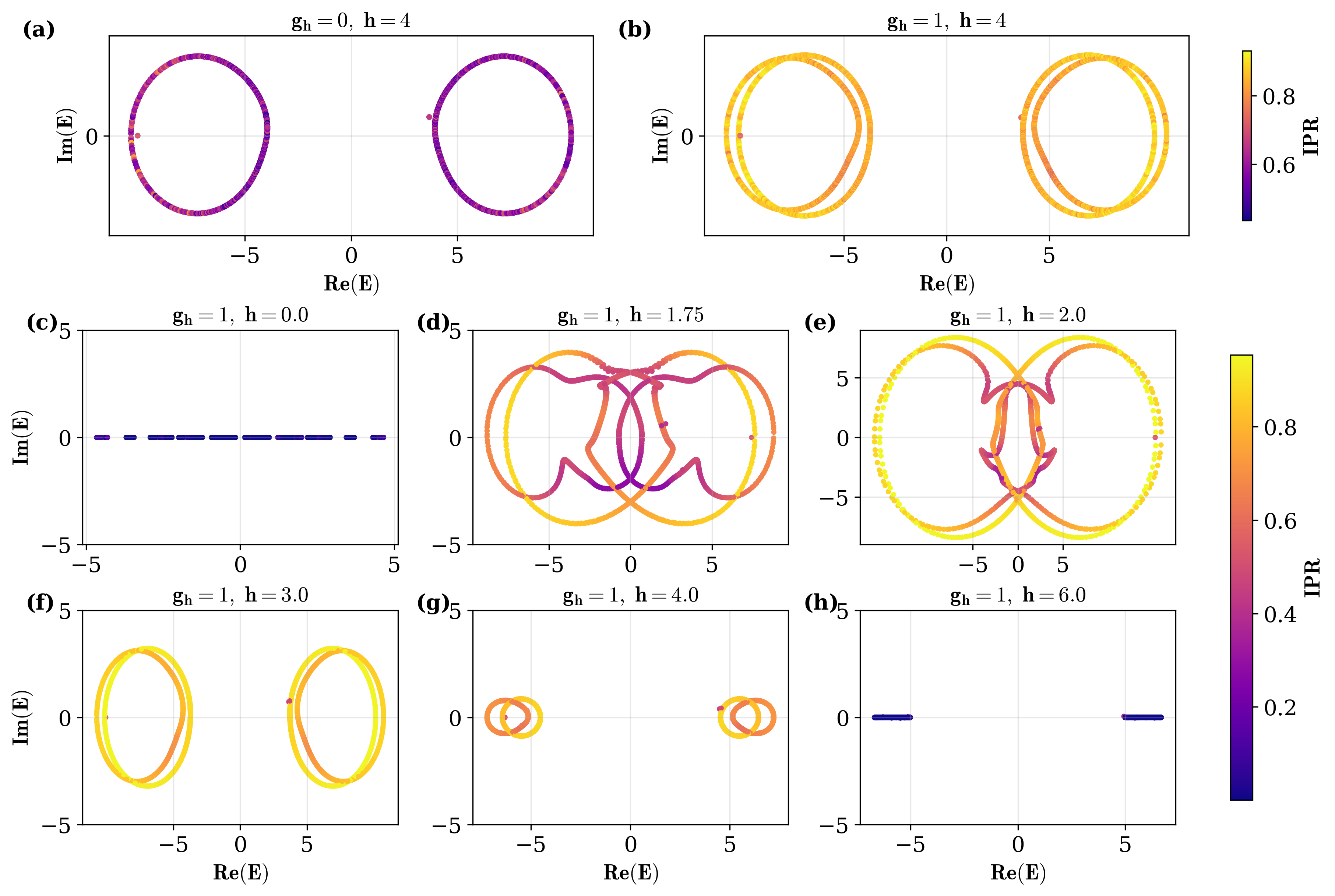}
		\caption{
			Complex-energy spectrum in the complex-energy plane colored by the inverse participation ratio (IPR). Panels (a,b) compare the spectra for $g_h=0$ and $g_h=1$ at fixed $h=4$, showing the splitting of spectral loops due to spin-dependent hopping. Panels (c--h) show the evolution of the spectrum with increasing non-Hermitian strength $h$ for $g_h=1$, demonstrating the real-to-complex-to-real spectral transition accompanied by localization and re-entrant delocalization behavior. L = 1597
		}
		\label{fig:complexspec}
	\end{figure*}	
	
	{\em Topological transition.—}
	
	We now characterize the different phases of the system from the viewpoint of spectral topology. 
	In non-Hermitian systems, complex eigenvalue spectra may form closed loops in the complex-energy plane, and these loops can be characterized by a spectral winding number~\cite{Longhi2019PRB,Gong2018,Liu2020}. 
	The appearance of finite spectral winding is closely correlated with the emergence of localized states and complex eigenvalues in quasiperiodic non-Hermitian systems~\cite{Padhan}. 
	In the present model, we find that the localization transition, the real-to-complex spectral transition, and the spectral-topological transition occur simultaneously over the same parameter regime.
	
	The winding number is defined as
	\begin{align}
		w = \lim_{L\rightarrow \infty}
		\frac{1}{2\pi i}
		\int_{0}^{2\pi}
		d\theta \,
		\partial_{\theta}
		\log
		\left[
		\det
		\left(
		H(\theta/L)-\varepsilon
		\right)
		\right],
		\label{eq:wind}
	\end{align}
	where $\varepsilon$ denotes the reference (base) energy. 
	Physically, the winding number counts the number of times the complex-energy spectrum encircles the base energy as the phase $\theta$ is varied from $0$ to $2\pi$~\cite{Longhi2019PRB,Padhan}. 
	For a purely real spectrum, the spectral loops collapse onto the real axis and the winding number vanishes, whereas finite winding numbers arise when the eigenvalues form closed contours in the complex-energy plane.
	
	In conventional non-Hermitian AAH models hosting a single localization transition, two base energies associated with the mobility-edge boundaries are generally sufficient to characterize the spectral topology~\cite{Padhan,Wu_2021,Liu2020}. 
	However, the present system exhibits a considerably richer structure due to the combined effect of quasiperiodicity, Rashba spin-orbit interaction, and spin-dependent hopping. 
	In particular, the system hosts two intermediate phases together with one fully localized phase, leading to multiple disconnected spectral contours in the complex-energy plane. 
	Consequently, a single global winding number becomes insufficient to fully characterize the spectral topology.
	
	To properly describe the topology of the disconnected spectral branches, we therefore define multiple winding numbers associated with different spectral contours. 
	Because the spectrum is symmetric under $\mathrm{Re}(E)\rightarrow -\mathrm{Re}(E)$, opposite contours contribute identical winding sectors. 
	Thus, although the full spectrum contains eight transition points associated with the spin-resolved mobility edges, only four independent winding numbers are required to characterize the topology. 
	These are denoted by $w_{1},w_{2},w_{3},w_{4}$.
	
	\begin{figure*}[t]
		\centering
		\includegraphics[width=0.98\textwidth]{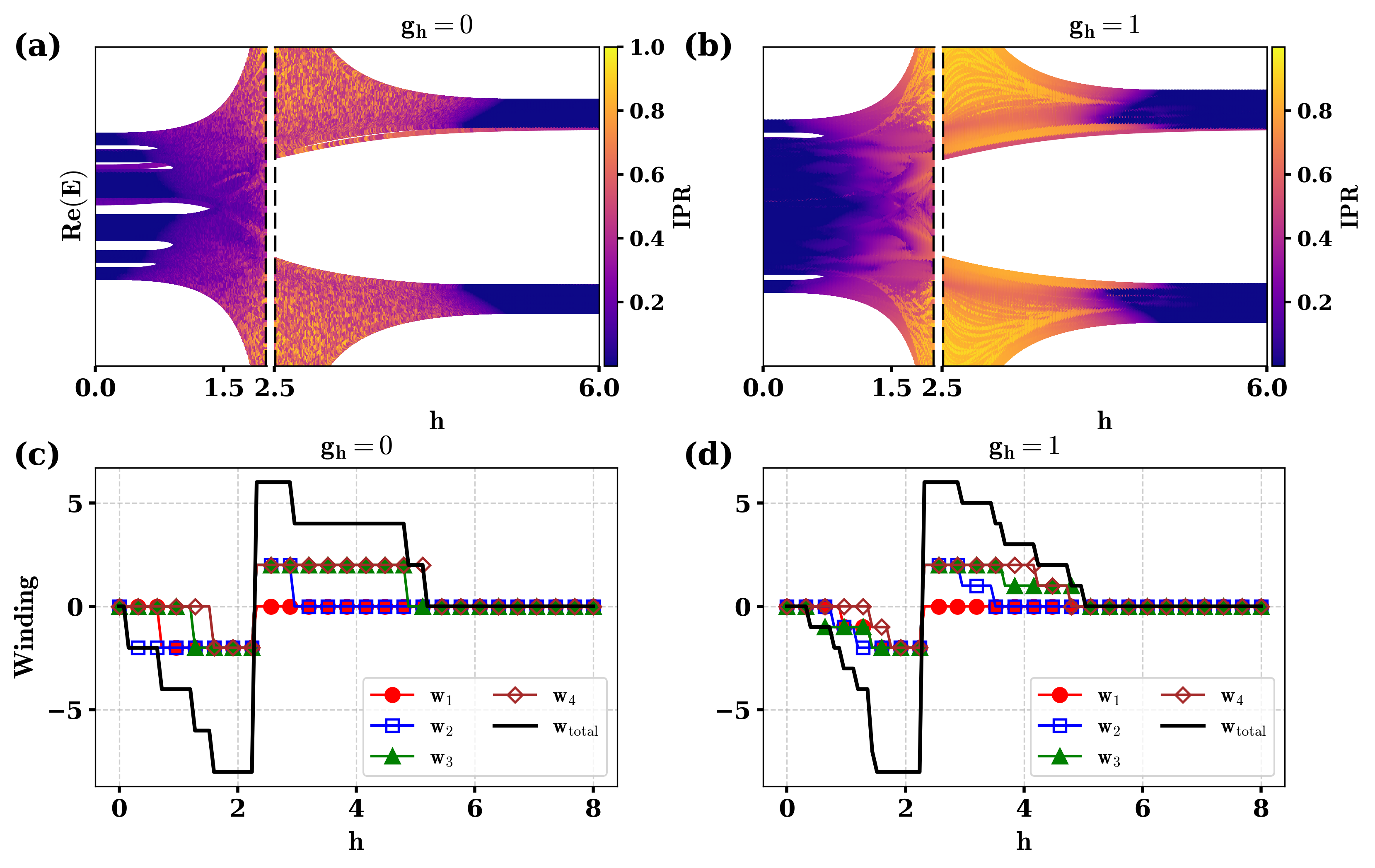}
		\caption{
			Upper panels: Evolution of the real part of the energy spectrum as a function of the non-Hermitian parameter $h$, colored by the inverse participation ratio (IPR), for (a) $g_h=0$ and (b) $g_h=1$.Here L = 610 The dashed vertical lines indicate the critical region associated with the localization transition. Lower panels: Corresponding winding numbers $w_1$, $w_2$, $w_3$, $w_4$, and total winding number $w_{\mathrm{tot}}$ as a function of $h$ for (c) $g_h=0$ and (d) $g_h=1$.L = 233.
				}
		\label{fig:winding_spectrum}
	\end{figure*}
	
	\subsection*{Effect of Spin-Dependent Hopping on Spectral Topology}
	
	Figure~\ref{fig:winding_spectrum} summarizes the evolution of the spectral winding numbers together with the localization properties of the system. 
	The upper panels display the real part of the spectrum as a function of the non-Hermitian parameter $h$, where the color scale denotes the inverse participation ratio (IPR). 
	The lower panels show the corresponding winding numbers associated with the disconnected spectral contours.
	
	For $g_h=0$, shown in Fig.~\ref{fig:winding_spectrum}(a,c), the system initially remains in an extended phase characterized by vanishing IPR and zero winding numbers. 
	As the non-Hermitian parameter increases, the eigenvalues gradually spread into the complex-energy plane and form two closed spectral loops. 
	These loops originate primarily from the two sublattice sectors of the SSH chain and remain nearly degenerate in spin space. 
	Simultaneously, the IPR increases significantly, indicating the onset of localization. 
	The emergence of finite winding numbers in this regime demonstrates that the appearance of localized states is strongly correlated with the formation of nontrivial spectral loops.
	
	The corresponding complex spectra are shown in Fig.~\ref{fig:complexspec}(a). 
	The two contours are approximately symmetric about the real axis and carry winding numbers $w=\pm2$, reflecting contributions from both spin sectors simultaneously. 
	Thus, in the absence of spin-dependent hopping, the topology remains effectively spin degenerate.
	
	A qualitatively different situation emerges when finite spin-dependent hopping $g_h$ is introduced. Most importantly, the spin-dependent hopping lifts the degeneracy between the spin-up and spin-down sectors and produces a complete restructuring of the spectral topology.
	
	The evolution of the corresponding complex spectra is shown in Fig.~\ref{fig:complexspec}(b--h). 
	For small non-Hermiticity ($h=0$), the spectrum remains entirely real and all winding numbers vanish, indicating a topologically trivial extended phase. 
	As $h$ increases, the eigenvalues spread into the complex plane and form multiple intertwined spectral contours. 
	Simultaneously, the IPR values become finite, signaling the onset of localization and mobility-edge-like behavior.
	
	The most remarkable feature appears at moderate non-Hermitian strength ($h\sim2$--$4$). 
	Unlike the $g_h=0$ case where the spectrum consisted of only two loops, each spectral contour now splits into two independent spin-resolved branches. 
	Consequently, the total number of spectral loops increases from two to four. 
	These four loops correspond to the spin-polarized channels $(\uparrow,\downarrow)$ associated with the two sublattice sectors of the SSH chain.
	
	This spectral splitting is a direct manifestation of the spin-dependent hopping structure. 
	The spin-up and spin-down channels acquire different effective complex hopping amplitudes, which modifies their spectral trajectories independently in the complex-energy plane. 
	As a result, the original spin-degenerate spectral loops separate into distinct spin-resolved spectral branches carrying distinct winding numbers.
	
	At intermediate values of $h$, the four contours become well separated and form two pairs of complex-conjugate loops. 
	The outer branches are associated with large IPR values, indicating strongly localized spin-polarized states, whereas the inner branches contain weakly localized or partially extended states. 
	This coexistence of strongly localized and weakly localized states reflects the mobility-edge-like structure already identified in the spin-resolved IPR analysis.
	
	The spectral splitting is also directly reflected in the winding-number distribution shown in Fig.~\ref{fig:winding_spectrum}(d). 
	For $g_h=0$, the spectral loops carry winding numbers $0$, $\pm2$, corresponding to the combined contribution from both spin sectors. 
	However, once spin degeneracy is lifted by finite $g_h$, the topology becomes considerably richer. 
	The separated spin-resolved loops now carry independent winding numbers such as $-1$, $0$, $1$, and $2$. 
	The appearance of these additional winding sectors demonstrates the formation of independent spin-resolved spectral branches carrying distinct winding numbers.
	
	To globally characterize the spectral topology, we define the total winding number as
	\begin{equation}
		w_{\mathrm{tot}}
		=
		w_1+w_2+w_3+w_4.
	\end{equation}
	The quantized plateaus of $w_{\mathrm{tot}}$ clearly distinguish the different topological phases appearing as the non-Hermitian parameter $h$ is varied.
	
	For larger values of $h$, the spectral loops gradually shrink and collapse back onto the real axis. 
	Simultaneously, the IPR decreases toward zero and all winding numbers vanish. 
	This indicates a re-entrant transition back to a topologically trivial extended phase with a predominantly real spectrum.
	
	The system therefore exhibits the re-entrant topological sequence
	\begin{equation}
		w=0
		\rightarrow
		w\neq0
		\rightarrow
		w=0,
	\end{equation}
	which occurs simultaneously with the localization transition
	\begin{equation}
		\mathrm{Extended}
		\rightarrow
		\mathrm{Localized}
		\rightarrow
		\mathrm{Extended}.
	\end{equation}
		
	The proposed non-Hermitian quasiperiodic SSH model with Rashba spin--orbit coupling can be experimentally realized using ultracold atoms loaded into optical lattices with Raman-assisted tunneling and synthetic spin--orbit coupling. A one-dimensional bichromatic optical lattice can be generated by superimposing two standing-wave laser fields with incommensurate wavelengths, thereby realizing the quasiperiodic Aubry--Andr\'e--Harper (AAH) modulation \cite{Roati2008,Luschen2018}. By introducing an additional staggered superlattice potential, each unit cell can be engineered to contain two inequivalent sites $A$ and $B$, producing the SSH-type dimerized structure with intracell and intercell hopping amplitudes \cite{Atala2013}.
	
	The non-Hermitian quasiperiodic onsite modulation can be implemented through laser-induced complex potentials or controlled dissipation schemes. In particular, spatially modulated gain and loss or complex light-induced Stark shifts can generate effective complex onsite terms of the form
	\[
	V_n=\frac{\lambda \cos(2\pi \beta n+\phi)}
	{1-\alpha \cos(2\pi \beta n+\phi)},
	\]
	where the phase $\phi=\theta+i h$ introduces the non-Hermitian parameter $h$.
	
	The spin degrees of freedom can be encoded in two hyperfine atomic states, while Rashba-type spin--orbit coupling can be generated using Raman-induced spin-momentum locking \cite{Lin2011,Galitski2013}. In this approach, Raman laser beams couple the two spin states while transferring momentum to the atoms, thereby inducing synthetic spin--orbit interactions analogous to Rashba coupling in condensed-matter systems. By suitably controlling the Raman beam phases and intensities, both spin-conserving and spin-flip hopping processes corresponding to the couplings $\alpha_{1}$, $\alpha_{2}$, $\alpha_{3}$, and $\alpha_{4}$ can be engineered.
	
	The spin-dependent hopping term $g_h$ can be realized through spin-selective Raman-assisted tunneling, where the effective hopping amplitudes for the two spin sectors acquire different complex phases. Importantly, although the hopping amplitudes contain imaginary components, the hopping Hamiltonian remains Hermitian because the forward and backward hopping processes are related by Hermitian conjugation. Such spin-dependent tunneling control and synthetic gauge-field engineering have already been demonstrated in ultracold-atom experiments using laser-assisted tunneling techniques \cite{Aidelsburger2013,Goldman2014}.
	
	In addition to ultracold atoms, the present model may also be implemented in photonic waveguide arrays, topolectric circuits, and other synthetic quantum platforms where complex onsite modulation and spin- or polarization-dependent couplings can be independently controlled.
	\section{Conclusion}
	
	In this work, we investigated the interplay between quasiperiodicity, Rashba spin--orbit coupling, and non-Hermitian modulation in a generalized quasiperiodic SSH lattice. By analyzing the localization properties, complex-energy spectra, and spectral winding numbers, we demonstrated the emergence of rich re-entrant localization and topological behavior driven by the non-Hermitian parameter $h$.
	
	Using the average inverse participation ratio (IPR) and spin-resolved IPR, we identified a sequence of extended, localized, and re-entrant extended phases as the non-Hermitian strength is increased. The system exhibits mobility-edge-like intermediate regimes where localized and extended states coexist within the same spectrum. The spin-resolved analysis further revealed that the localization properties of the spin-up and spin-down sectors evolve differently in the presence of spin-dependent hopping, leading to spin-selective localization patterns and distinct spectral structures.
	
	A central result of the present work is the direct correspondence between localization and spectral topology in the complex-energy plane. In the absence of spin-dependent hopping, the spectrum forms two nearly spin-degenerate loops characterized by winding numbers $w=\pm2$. Upon introducing finite spin-dependent hopping, the spin degeneracy is lifted and each spectral contour splits into two independent spin-resolved branches. Consequently, the number of disconnected spectral loops increases from two to four, producing multiple winding sectors associated with different spectral branches. The total winding number exhibits quantized plateaus that clearly distinguish the various topological regimes of the system.
	
	We further showed that the formation of complex spectral loops is accompanied by enhanced localization and finite winding numbers, whereas the collapse of the loops back onto the real axis restores the extended phase together with a topologically trivial spectral structure. This demonstrates a re-entrant transition sequence involving simultaneous localization-delocalization and spectral-topological transitions. The emergence of multiple disconnected spectral branches and spin-resolved winding sectors highlights the nontrivial role played by spin-dependent hopping in reshaping the topology of non-Hermitian quasiperiodic systems.
	
	Overall, our results reveal a rich interplay between quasiperiodicity, spin--orbit interaction, and non-Hermitian modulation, leading to unconventional spectral topology and spin-selective localization phenomena. The present model provides a possible platform for realizing controllable non-Hermitian topological phases and spin-resolved spectral engineering in photonic lattices, cold-atom systems, topolectric circuits, and other synthetic quantum platforms.	
	
	\bibliography{Bib.bib}

\end{document}